\documentclass[11pt]{article}
\usepackage[utf8]{inputenc}

\usepackage[paperwidth=8.5in,paperheight=11in,portrait,top=1in,bottom=1.in,left=1.15in,right=1.15in]{geometry}
\usepackage{authblk}

\usepackage{amsfonts,amsmath,amssymb,amsthm}
\allowdisplaybreaks
\usepackage{latexsym,mathrsfs,mathtools,bm}
\usepackage{braket}
\usepackage{pgfplots}
\usepackage{graphicx,subcaption,epsfig,caption,float,xcolor}
\usepackage{enumitem}

\usepackage[hidelinks]{hyperref}
\usepackage{bookmark}

\theoremstyle{plain}
\newtheorem{thm}{Theorem}[section]

\newtheorem{rem}[thm]{Remark}

\numberwithin{equation}{section}

\def\cA{{\mathcal A}}      \def\cC{{\mathcal C}}
      
   \def\cH{{\mathcal H}}   
      
   \def\cN{{\mathcal N}}   \def\cO{{\mathcal O}}

\title{\bf A discrete Smorodinsky--Winternitz II superintegrable system}

\renewcommand*{\Affilfont}{\normalsize\small}
\author[1]{Pierre-Antoine Bernard}
\author[2]{Vutha Vichhea Chea}
\author[3]{Luc Vinet}
\affil[1]{Department of Computer Science, University of Toronto, Toronto, ON, Canada}
\affil[2,3]{Centre de Recherches Math\'ematiques, Universit\'e de Montr\'eal, P.O. Box 6128, Centre-ville Station, Montr\'eal (Qu\'ebec), H3C 3J7, Canada. \vspace{.5em}}

{
	\makeatletter
	\renewcommand\AB@affilsepx{: \protect\Affilfont}
	\makeatother
	\affil[ ]{E-mail addresses}
	\makeatletter
	\renewcommand\AB@affilsepx{, \protect\Affilfont}
	\makeatother
    \affil[1]{bernardpierreantoine@outlook.com}
	\affil[2]{vutha.vichhea.chea@umontreal.ca}
	\affil[3]{luc.vinet@umontreal.ca}
}

\begin{document}

\maketitle

\begin{abstract}

We construct a discrete Smorodinsky--Winternitz II superintegrable system on a triangular region of the two-dimensional square lattice. The model is built from a pair of commuting finite-difference number operators with finite spectrum together with an associated ladder-operator structure. We show that it is maximally superintegrable and that its symmetry algebra admits a Hahn-algebra presentation. The spectral problem is solved exactly in terms of bivariate orthogonal polynomials of mixed Krawtchouk and dual Hahn type associated with the factorized $A_2$-Leonard pair. Finally, we show that the continuum limit recovers the continuous Smorodinsky--Winternitz II system together with its Hermite--Laguerre eigenfunctions. We further explain how the Hahn presentation of the discrete symmetry algebra becomes singular in this limit, while the limiting algebraic structure is naturally described by the Laguerre--Heun algebra associated with Cartesian and parabolic separation of variables.

\end{abstract}

\section{Introduction}

The Smorodinsky--Winternitz II system is one of the two-dimensional potentials singled out by
Fri\v{s}, Mandrosov, Smorodinsky, Uhl\'i\v{r} and Winternitz as admitting two independent
second-order integrals of motion \cite{frivs1965higher,winternitz1966symmetry}. Physically it is a
$2\!:\!1$ anisotropic oscillator carrying a centrifugal barrier along one axis. Its two integrals
are the separation constants of the Cartesian and of the parabolic coordinate systems, and their
commutator closes on a polynomial algebra. We construct here a finite model that carries the same
structure on a triangular region of the two-dimensional square lattice: a pair of commuting
finite-difference number operators with finite spectrum, a ladder-operator algebra built on them,
three integrals of motion, and a spectral problem solved in closed form.

The eigenfunctions are bivariate polynomials of Tratnik type \cite{tratnik1991some} combining a
Krawtchouk and a dual Hahn factor, the family attached to the factorized $A_2$-Leonard pair
\cite{Cra-Zai-25}. The symmetry algebra is cubic and admits a Hahn-algebra presentation
\cite{granovskii1992mutual,Fra-Gab-Vin-Vin-Zhe-19}, which is what accounts for the dual Hahn factor.

The continuum limit is where the two settings part company. The Hamiltonian, the ladder operators
and the eigenfunctions all converge to their continuous counterparts, so the finite model is a
genuine realization of the Smorodinsky--Winternitz II system and not a finite-difference
approximation of its equations of motion. The Hahn presentation, however, does not survive: the
redefinition that brings the cubic algebra to Hahn form is tied to the lattice normalization and
diverges with the lattice size. The symmetry operators themselves have regular limits, and they
close instead on the Laguerre--Heun algebra \cite{chea20262d} associated with the Cartesian and
parabolic separation of variables. The finite model thus supplies a discrete origin for that
algebra.

This paper is the second of a pair. The companion paper \cite{Che-Vin-26} treats the
Smorodinsky--Winternitz I system, and we refer to it, and to the review \cite{Mil-Pos-Win-13}, for
the setting common to both: the notion of superintegrability used throughout, the polynomial
symmetry algebras of the two-dimensional systems and their ties to special functions
\cite{Das-01,Mar-09,Tem-Tur-Win-00}, the Racah-type algebras that organize them
\cite{genest2014superintegrability}, the representation-theoretic perspective
\cite{reshetikhin2015degenerately,arthamonov2021superintegrable}, the difference equations arising
in the study of multivariate orthogonal polynomials whose degeneracies stem from symmetries
\cite{Ata-Pog-Vic-Wol-01,Ata-Nat-Pog-Geo-Vic-Wol-01,de2017superintegrable,Gen-Vin-14,iliev2018symmetry,Ili-Xu-20,kalnins2011two,Sas-23},
and the finite discrete oscillators
\cite{Gab-Gen-Lem-Vin-15,Gen-Mik-Vin-Yu-17,Mik-Pos-Vin-Zhe-2012} --- in particular the
two-dimensional model of \cite{Mik-Pos-Vin-Zhe-2012}, built on the bivariate Krawtchouk polynomials
of Griffiths type \cite{Gri-71}, from which the present program starts. Apart from that background,
what follows is self-contained.

Section~\ref{sec:model-definition} sets up the lattice, the weight and the Hamiltonian.
Section~\ref{sec:ladder} builds the ladder operators, establishes maximal superintegrability,
identifies the symmetry algebra and solves the spectral problem. The continuum limit occupies
Section~\ref{sec:continuum-limit}: first the Hamiltonian and its eigenfunctions, then the symmetry
algebra and the degeneration of its Hahn presentation. Section~\ref{sec:conclusion} closes with the results and concluding remarks. The
coefficients of the ladder operators are gathered in Appendix~\ref{app:ladder-coefficients}, and
the special-function formulas used throughout in Appendices~\ref{app:hyp-poly}
and~\ref{app:Tratnik-poly}.

\section{A discrete Smorodinsky--Winternitz II model}
\label{sec:model-definition}

\begin{figure}[htbp]
\centering
\begin{tikzpicture}[scale=1.1]

\def\N{5}

\draw[gray!30, step=1] (0,0) grid (\N+0.5,\N+0.5);

\draw[->, thick] (0,0) -- (\N+0.8,0) node[right] {$x_1$};
\draw[->, thick] (0,0) -- (0,\N+0.8) node[above] {$x_2$};

\draw (0,0) -- (0,-0.1) node[below] {$0$};
\draw (0,0) -- (-0.1,0) node[left] {$0$};

\draw (\N,0) -- (\N,-0.1) node[below] {$N$};
\draw (0,\N) -- (-0.1,\N) node[left] {$N$};

\foreach \x in {0,...,\N}
{
    \pgfmathtruncatemacro{\ymax}{\N-\x}
    \foreach \y in {0,...,\ymax}
    {
        \fill (\x,\y) circle (2pt);
    }
}

\end{tikzpicture}
\caption{The discrete position space $\mathfrak{R}(N)$ of \eqref{eq:triangular-region},
an isosceles right triangular region of side length $N$ in the first quadrant of the
$(x_1,x_2)$-plane (here $N=5$).}
\label{fig:triangular-lattice-II}
\end{figure}
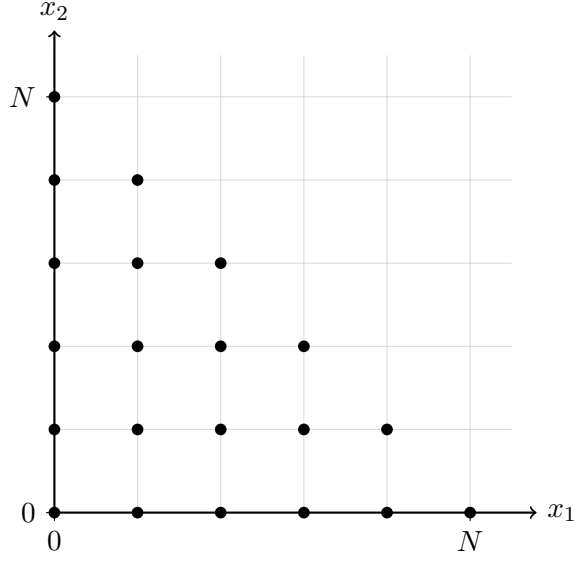

We now construct the discrete Smorodinsky--Winternitz II model. The
position space is a triangular region of the two-dimensional square lattice whose geometry is adapted to the mixed Krawtchouk and dual Hahn structure that will appear in the spectral problem. More precisely, we take the position space to be the region (see Figure~\ref{fig:triangular-lattice-II})
\begin{equation}
    \mathfrak{R}(N)
    = \bigl\{ (x_1,x_2) \in \mathbb{N}_0^2 \;\big|\; 0 \le x_1 + x_2 \le N \bigr\},
    \qquad N \in \mathbb{N},
    \label{eq:triangular-region}
\end{equation}
The associated Hilbert space is generated by the position eigenstates
localized on this region,
\begin{equation}
    \mathfrak{H} = \text{Span}\{\ket{x_1,x_2} | \, (x_1,x_2) \in \mathfrak{R}(N) \}.
\end{equation}
The scalar product is chosen so that the position basis satisfies
\begin{equation}
    \braket{x_1,x_2|y_1,y_2} =
    \frac{\delta_{x_1,y_1}\delta_{x_2,y_2}}{w(x_1,x_2)},
    \label{eq:ortho-rel-position}
\end{equation}
where \(\delta_{x_i,y_i}\) is the Kronecker delta and
\(w(x_1,x_2)\) is a positive local weight function to be determined
below. With this convention, the completeness relation is
\begin{equation}
    \sum_{(x_1,x_2) \in \mathfrak{R}(N)}
    w(x_1,x_2)\ket{x_1,x_2}\bra{x_1,x_2} = 1,
    \label{eq:completeness-relation}
\end{equation}
so that every vector state \(\ket{\psi}\in\mathfrak{H}\) admits the expansion
\begin{equation}
    \ket{\psi} =
    \sum_{(x_1,x_2) \in \mathfrak{R}(N)}
    w(x_1,x_2)\psi(x_1,x_2)\ket{x_1,x_2},
    \qquad
    \braket{x_1,x_2|\psi} = \psi(x_1,x_2).
    \label{eq:state-expansion}
\end{equation}
Any operator acting on \(\mathfrak{H}\) is represented in the position basis by
\begin{equation}
    \cO\ket{\psi}
    =
    \sum_{(x_1,x_2) \in \mathfrak{R}(N)}
    w(x_1,x_2)\cO\psi(x_1,x_2)\ket{x_1,x_2},
    \qquad
    \cO\psi(x_1,x_2) = \bra{x_1,x_2}\cO\ket{\psi}.
    \label{eq:linear-operator-action}
\end{equation}
Hermitian conjugation is taken with respect to the scalar product above,
namely
\begin{equation}
    \sum_{(x_1,x_2) \in \mathfrak{R}(N)}
    w(x_1,x_2)\phi^*(x_1,x_2)\cO\psi(x_1,x_2)
    =
    \sum_{(x_1,x_2) \in \mathfrak{R}(N)}
    w(x_1,x_2)\cO^\dagger\phi^*(x_1,x_2)\psi(x_1,x_2).
    \label{eq:adjoint-rule}
\end{equation}

\subsection{Hamiltonian operator}

The Hamiltonian of the discrete model is taken to be the sum of two
commuting number operators,
\begin{equation}
    H = N_1 + N_2 + \frac{\alpha}{2} + 1,
    \qquad [N_1 , N_2] = 0.
    \label{eq:hamiltonian-operator}
\end{equation}
Here \(\alpha > -1\) is a fixed parameter. The explicit expressions
for \(N_1\) and \(N_2\) are chosen so that the operators preserve the
space of functions on \(\mathfrak{R}(N)\), are self-adjoint with respect to a local
positive weight function, and admit a common family of polynomial
eigenfunctions. They are given by
\begin{align}
    N_1 &= - \Biggl[
    \frac{p(N-x_1-x_2)(\alpha+2N-x_1-x_2+1)}{(\alpha+2N-x_1-2x_2+1)}\,\Delta_{1,0}
    + \frac{p\,x_2(\alpha+N-x_2+1)}{(\alpha+2N-x_1-2x_2+1)}\,\Delta_{1,-1}
    \nonumber\\
    &\qquad\quad
    + \frac{(1-p)x_1(N-x_2)}{(\alpha+2N-x_1-2x_2+1)}\,\Delta_{-1,1}
    + \frac{(1-p)x_1(\alpha+N-x_1-x_2+1)}{(\alpha+2N-x_1-2x_2+1)}\,\Delta_{-1,0}
    + p\,N_2
    \Biggr],
    \nonumber\\
    N_2 &= - \Biggl[
    \frac{(N-x_2)(N-x_1-x_2)(\alpha+2N-x_1-x_2+1)}{(\alpha+2N-x_1-2x_2)_2}\,\Delta_{0,1}
    \nonumber\\
    &\qquad\quad\
    + \frac{x_2(\alpha+N-x_2+1)(\alpha+N-x_1-x_2+1)}{(\alpha+2N-x_1-2x_2+1)_2}\,\Delta_{0,-1}
    \Biggr],
    \label{eq:number-operators}
\end{align}
where \(p\in(0,1)\) is another fixed parameter,
\((a)_k=a(a+1)\cdots(a+k-1)\) is the Pochhammer symbol, and
\(\Delta_{\eta_1,\eta_2}\), where $(\eta_1,\eta_2)\in\mathbb{Z}^2$, denotes a finite difference operator whose action on a function $f$, reads as
\begin{align}
    \Delta_{\eta_1,\eta_2}f(x_1,x_2)
    =
    \begin{cases}
        f(x_1+\eta_1, x_2+\eta_2) - f(x_1,x_2) \qquad \text{if} \qquad (\eta_1,\eta_2) \ne (0,0),
        \\
        f(x_1,x_2) \qquad \text{if} \qquad (\eta_1,\eta_2) = (0,0).
    \end{cases}
    \label{eq:difference-operator}
\end{align}
\begin{rem}
    The parameter $p$ is specific to the discrete model. The continuum limit of Section~\ref{sec:continuum-limit} requires instead that $p$ tend to zero with $N$, and its effect disappears in that respect.
\end{rem}
The weight function \(w(x_1,x_2)\) is obtained by imposing the
Hermiticity conditions
\begin{equation}
    N_i^\dagger = N_i, \qquad i=1,2.
\end{equation}
A direct computation using \eqref{eq:adjoint-rule} yields the
factorized form
\begin{align}
    &w(x_1,x_2) = w_1(x_1)w_2(x_1,x_2), \qquad w_1(x_1) = \binom{N}{x_1}p^{x_1}(1-p)^{N-x_1},
    \nonumber\\
    &w_2(x_1,x_2) = \frac{\alpha+2N-x_1-2x_2+1}{\alpha+2N-x_1-x_2+1}\frac{\frac{\Gamma(\alpha+N-x_2+1)}{\Gamma(N-x_2+1)}\frac{\Gamma(\alpha+N-x_1-x_2+1)}{\Gamma(N-x_1-x_2+1)}}{\frac{\Gamma(\alpha+2N-x_1-x_2+1)}{\Gamma(2N-x_1-x_2+1)}}
    \frac{\binom{2N-x_1}{x_2}}{\binom{2N-x_1}{N}}.
    \label{eq:weight-function}
\end{align}
\begin{rem}
The restrictions $\alpha>-1$ and $p\in(0,1)$
guarantee that the weight function is strictly positive
throughout $\mathfrak{R}(N)$.
\end{rem}
The Hamiltonian \eqref{eq:hamiltonian-operator}, together with the
domain \eqref{eq:triangular-region} and the weight
\eqref{eq:weight-function}, defines the discrete
Smorodinsky--Winternitz II model. In
Section~\ref{sec:continuum-limit}, we show that an appropriate scaling
of the lattice variables, followed by a gauge transformation, yields
its continuous counterpart.

\section{Ladder operators, superintegrability and exact solvability}
\label{sec:ladder}

\subsection{Dynamical algebra}
\label{subsec:dynamical-algebra}

We assume the existence of two pairs of ladder operators $\{a_i, a_i^\dagger\}$ through which their defining action is specified through the commutation relations
\begin{equation}
    [N_i,a_j] = -\delta_{ij}\,a_j, \qquad [N_i,a_j^\dagger] = \delta_{ij}\,a_j^\dagger.
    \label{eq:Ni-aj-commutation}
\end{equation}
These relations ensure that the operators $a_i$ and $a_i^\dagger$ lower and raise, respectively, the eigenvalues of $N_i$ by one unit. This ladder structure provides a natural starting point for the construction of the dynamical algebra of the model. We also assume that each annihilation operator $a_i$ can be expressed as a linear combination of finite-difference operators,
\begin{equation}
    a_i=
    \sum_{|\eta_1|+|\eta_2|\le 2N}
    a_{i}^{\eta_1,\eta_2}\,
    \Delta_{\eta_1,\eta_2},
    \label{eq:ak-ansatz}
\end{equation}
where the coefficients $a_i^{\eta_1,\eta_2}$ are functions on $\mathfrak{R}(N)$ that vanish outside the region. Using the explicit representations of the number operators~\eqref{eq:number-operators},  the commutation relations~\eqref{eq:Ni-aj-commutation} translate into an inhomogeneous linear system for the coefficients $a_i^{\eta_1,\eta_2}$ whose solution is subjected to the confinement on $\mathfrak{R}(N)$, to the lowest admissible order of finite-differences and to a normalization constant. One
obtains
\begin{align}
    a_1
    &=
    \sqrt{\frac{1-p}{pN}}
    \sum_{\vec{\eta}\in S_1}
    a_{1}^{\eta_1,\eta_2}\,
    \Delta_{\eta_1,\eta_2},
    \nonumber\\
    a_2
    &=
    \frac{\sqrt{1-p}}{N}
    \sum_{\vec{\eta}\in S_2}
    a_{2}^{\eta_1,\eta_2}\,
    \Delta_{\eta_1,\eta_2},
    \label{eq:explicit-annihilation}
\end{align}
while Hermitian conjugation with respect to the scalar product
\eqref{eq:adjoint-rule} yields the corresponding creation operators
\begin{align}
    a_1^\dagger
    &=
    \sqrt{\frac{1-p}{pN}}
    \sum_{\vec{\eta}\in\bar S_1}
    \bar a_{1}^{\eta_1,\eta_2}\,
    \Delta_{\eta_1,\eta_2},
    \nonumber\\
    a_2^\dagger
    &=
    \frac{\sqrt{1-p}}{N}
    \sum_{\vec{\eta}\in\bar S_2}
    \bar a_{2}^{\eta_1,\eta_2}\,
    \Delta_{\eta_1,\eta_2}.
    \label{eq:explicit-creation}
\end{align}
For readability, the explicit coefficient functions are collected in
Appendix~\ref{app:ladder-coefficients}. Their algebraic properties,
however, can be expressed in closed form through the structure
functions
\begin{align}
    &a_1^\dagger a_1
    =
    \frac{(N-N_1-N_2+1)N_1}{N},
    \qquad
    a_1a_1^\dagger
    =
    \frac{(N-N_1-N_2)(N_1+1)}{N},
    \nonumber\\
    &a_2^\dagger a_2
    =
    \frac{(N-N_1-N_2+1)(N-N_2+1)(N_2+\alpha)N_2}{N^2},
    \nonumber\\
    &a_2a_2^\dagger
    =
    \frac{(N-N_1-N_2)(N-N_2)(N_2+\alpha+1)(N_2+1)}{N^2},
    \nonumber\\
    &(N-N_1-N_2)a_i^\dagger a_j
    =
    (N-N_1-N_2+1)a_ja_i^\dagger,
    \qquad i\neq j.
    \label{eq:structure-function}
\end{align}
The complete algebraic structure of the \emph{dynamical algebra} can thus be summarized as
\begin{align}
    &[N_i,N_j]=[a_i,a_j]=0,
    \qquad
    [N_i,a_j]=-\delta_{ij}a_j,
    \qquad
    [a_i,a_i^\dagger]
    =
    a_ia_i^\dagger-a_i^\dagger a_i,
    \nonumber\\
    &(N-N_1-N_2+1)[a_i,a_j^\dagger]
    =
    -a_j^\dagger a_i,
    \qquad i\neq j.
    \label{eq:dynamical-algebra}
\end{align}
These relations provide the algebraic framework underlying the discrete
model and will be used in the next subsections to construct its
symmetry algebra and solve the spectral problem.

\subsection{Symmetry algebra}

We now exhibit the symmetries of the Hamiltonian~\eqref{eq:hamiltonian-operator}. In accordance with the continuous Smorodinsky--Winternitz II model as will be shown in Section \ref{sec:continuum-limit}, we consider the following algebraically independent operators,
\begin{equation}
    C_1
    =
    N_2+\frac{\alpha+1}{2},
    \qquad
    C_2
    =
    \{a_1,a_2^\dagger\}
    +
    \{a_1^\dagger,a_2\},
    \label{eq:symmetry-operators}
\end{equation}
where $\{\cdot,\cdot\}$ denotes the anticommutator. Using the
commutation relations of the dynamical algebra
\eqref{eq:dynamical-algebra}, one immediately verifies that
\begin{equation}
    [H,C_i]=0,
    \qquad i=1,2.
\end{equation}
Hence $C_1$ and $C_2$ are symmetry operators of the Hamiltonian.
Together with $H$, the discrete Smorodinsky--Winternitz II model admits three integrals of motion in two dimensions, establishing it to be \emph{maximally superintegrable}.

A direct computation further shows that the symmetry operators close
under commutation and satisfy the cubic algebra
\begin{align}
    &[C_1,C_2]\equiv C_3,
    \nonumber\\
    &[C_2,C_3]
      =
      \mu C_1^3
      +
      \nu C_1^2
      +
      \xi C_1
      +
      \zeta,
    \nonumber\\
    &[C_3,C_1]
      =
      -C_2.
\end{align}
The constants which characterize the corresponding algebra are given by
\begin{align}
    &\mu
    =
    8\eta,
    \qquad
    \nu
    =
    -3\eta\bigl(2(N+H+1)+\alpha\bigr),
    \qquad
    \xi
    =
    \eta\Bigl(2(2N+\alpha+2)H
    +
    1
    +
    (1-\alpha^2)\Bigr),
    \nonumber\\
    &\zeta
    =
    -\frac14
    \eta
    (1-\alpha^2)
    \bigl(2(N+H+1)+\alpha\bigr),
    \qquad
    \eta
    =
    \frac{\bigl(2(N-H+1)+\alpha+1\bigr)^2}{N^3}.
    \label{eq:struc-const}
\end{align}
\begin{rem}
The quantities $\mu,\nu,\xi,\zeta,\eta$ are structure
functions rather than scalar constants: each is central, being a
function of the Hamiltonian $H$, and therefore commutes with all
the generators $C_i$.
\end{rem}
We shall refer to the algebra generated by
$\{C_1,C_2,C_3,H\}$ as the discrete two-dimensional
Smorodinsky--Winternitz II algebra,
denoted by $\mathcal{SW}^{N}_{II}(2)$.
It admits the Casimir operator of the generalized Daskaloyannis type \cite{Das-91,Das-01,Mar-09}
\begin{equation}
    C
    =
    \frac{\mu}{2}C_1^4
    +
    \frac{2\nu}{3}C_1^3
    +
    \left(\frac{\mu}{2}+\xi\right)C_1^2
    +
    \left(\frac{\nu}{3}+2\zeta\right)C_1
    -
    C_2^2
    +
    C_3^2,
\end{equation}
which commutes with all generators. In the present realization, the
Casimir reduces to
\begin{equation}
    C
    =
    -\frac14
    \eta
    (1-\alpha^2)
    \left(2(2N+\alpha+2)H+1\right).
\end{equation}

Although the symmetry algebra naturally appears in the cubic form
above, it can be brought to the standard Hahn-algebra presentation.
Indeed, introducing
\begin{equation}
    \widetilde C_1=C_1,
    \qquad
    \widetilde C_2
    =
    C_2+\kappa C_1^2+\lambda C_1,
    \label{eq:hahn-pres}
\end{equation}
with
\begin{equation}
    \kappa
    =
    \sqrt{\frac{\mu}{2}},
    \qquad
    \lambda
    =
    \frac{\nu}{3\kappa},
\end{equation}
one obtains
\begin{align}
    &[\widetilde C_1,\widetilde C_2]
      \equiv
      \widetilde C_3,
    \nonumber\\
    &[\widetilde C_2,\widetilde C_3]
      =
      \kappa
      \{\widetilde C_1,\widetilde C_2\}
      +
      \lambda\widetilde C_2
      +
      (\xi-\lambda^2)\widetilde C_1
      +
      \zeta,
    \nonumber\\
    &[\widetilde C_3,\widetilde C_1]
      =
      \kappa\widetilde C_1^{\,2}
      +
      \lambda\widetilde C_1
      -
      \widetilde C_2,
      \label{eq:hahn-algebra}
\end{align}
which is precisely the Hahn algebra \cite{Fra-Gab-Vin-Vin-Zhe-19,granovskii1992mutual}. Its emergence therefore provides an algebraic explanation for the appearance of dual Hahn polynomials in the spectral theory of the model. In this sense, the symmetry algebra and the exact eigenfunctions constitute two complementary manifestations of the same underlying structure.

\subsection{Exact solution}

Having shown the model to be maximally superintegrable, we now turn to its exact solvability, in agreement with the conjecture proposed by Tempesta, Turbiner and Winternitz \cite{Tem-Tur-Win-00}.
Since the number operators commute, the spectral problem can be solved
simultaneously. We therefore introduce a complete orthonormal basis of
energy eigenstates
$\{\ket{n_1,n_2}\}$ satisfying
\begin{align}
    &N_i\ket{n_1,n_2}
      =
      n_i\ket{n_1,n_2},
    \label{eq:number-action}
    \\
    &\braket{m_1,m_2|n_1,n_2}
      =
      \delta_{m_1,n_1}
      \delta_{m_2,n_2}.
    \label{eq:ortho-rel-energy-ket}
\end{align}
The structure functions
\eqref{eq:structure-function}
immediately imply that the energy spectrum is finite. Indeed,
\begin{align}
    \begin{rcases}
    a_i\ket{n_1,n_2}=0
    \qquad\text{if}\qquad n_i=0
    \nonumber\\
    a_i^\dagger\ket{n_1,n_2}=0
    \qquad\text{if}\qquad n_1+n_2=N
    \end{rcases}
    \qquad
    \Longrightarrow
    \qquad
    (n_1,n_2)\in\mathfrak{R}(N).
\end{align}
The action of the ladder operators is therefore
\begin{align}
    &a_1\ket{n_1,n_2}
      =
      \sqrt{\frac{n_1(N-n_1-n_2+1)}{N}}
      \ket{n_1-1,n_2},
    \nonumber\\
    &a_1^\dagger\ket{n_1,n_2}
      =
      \sqrt{\frac{(n_1+1)(N-n_1-n_2)}{N}}
      \ket{n_1+1,n_2},
    \nonumber\\
    &a_2\ket{n_1,n_2}
      =
      \sqrt{\frac{n_2(n_2+\alpha)(N-n_2+1)(N-n_1-n_2+1)}{N^2}}
      \ket{n_1,n_2-1},
    \nonumber\\
    &a_2^\dagger\ket{n_1,n_2}
      =
      \sqrt{\frac{(n_2+1)(n_2+\alpha+1)(N-n_2)(N-n_1-n_2)}{N^2}}
      \ket{n_1,n_2+1},
    \label{eq:ladder-action}
\end{align}
where the normalization constants have been chosen to be real and positive. Expanding the energy eigenstates in the position basis \eqref{eq:state-expansion},
\begin{equation}
    \ket{n_1,n_2}
    =
    \sum_{(x_1,x_2)\in\mathfrak{R}(N)}
    w(x_1,x_2)
    P_{n_1,n_2}(x_1,x_2)
    \ket{x_1,x_2},
    \qquad
    \braket{x_1,x_2|n_1,n_2}
    =
    P_{n_1,n_2}(x_1,x_2),
\end{equation}
the corresponding orthogonality relation follows from \eqref{eq:ortho-rel-energy-ket} and \eqref{eq:ortho-rel-position},
\begin{equation}
    \sum_{x_1=0}^{N}
    w_1(x_1)
    \sum_{x_2=0}^{N-x_1}
    w_2(x_1,x_2)
    P_{m_1,m_2}^*(x_1,x_2)
    P_{n_1,n_2}(x_1,x_2)
    =
    \delta_{m_1,n_1}\delta_{m_2,n_2},
    \label{eq:ortho-rel-energy-fct}
\end{equation}
Likewise, the eigenvalue equations
\eqref{eq:number-action}, together with the explicit realizations of
the number operators \eqref{eq:number-operators}, lead to the pair of difference equations
\begin{equation}
    N_iP_{n_1,n_2}(x_1,x_2) = n_iP_{n_1,n_2}(x_1,x_2), \qquad i = 1,2.
    \label{eq:diff-eq-energy}
\end{equation}
The spectral problem defined by \eqref{eq:ortho-rel-energy-fct} and \eqref{eq:diff-eq-energy} coincides with that studied in \cite{Cra-Zai-25}. It follows that the eigenfunctions are given by the bivariate Krawtchouk and dual Hahn polynomials of Tratnik type \cite{tratnik1991some} (see Appendix~\ref{app:Tratnik-poly}),
\begin{align}
    &P_{n_1,n_2}(x_1,x_2)
    =
    K_{n_1,n_2}(x_1;p,N)
    R_{n_2}(x_1,x_2;\alpha,N),
    \label{eq:energy-eigenfct}
    \\
    &K_{n_1,n_2}(x_1;p,N)
    =
    (-1)^{n_1}
    (1-p)^{-n_2/2}
    \sqrt{
    \binom{N-n_2}{n_1}
    \left(\frac{p}{1-p}\right)^{n_1}}
    \,
    \hat K_{n_1}(x_1;p,N-n_2),
    \label{eq:mod-Krawtchouk}
    \\
    &R_{n_2}(x_1,x_2;\alpha,N)
    =
    (-1)^{n_2}
    \binom{N-x_1}{n_2}
    \sqrt{\frac{n_2!}{\Gamma(n_2+\alpha+1)}}
    \,
    \hat R_{n_2}
    \!\left(
    \lambda(x_2);
    -N-1,
    x_1-N-\alpha-1,
    N-x_1
    \right),
    \label{eq:mod-dual-Hahn}
\end{align}
where $\hat K_{n_1}$ and $\hat R_{n_2}$ denote the Krawtchouk and dual Hahn polynomials introduced in Appendix~\ref{app:hyp-poly}. The energy eigenfunctions therefore satisfy
\begin{align}
    &HP_{n_1,n_2}
    =
    \left(
    n_1+n_2+\frac{\alpha}{2}+1
    \right)
    P_{n_1,n_2},
    \nonumber\\
    &a_1P_{n_1,n_2}
    =
    \sqrt{\frac{n_1(N-n_1-n_2+1)}{N}}
    P_{n_1-1,n_2},
    \nonumber\\
    &a_1^\dagger P_{n_1,n_2}
    =
    \sqrt{\frac{(n_1+1)(N-n_1-n_2)}{N}}
    P_{n_1+1,n_2},
    \nonumber\\
    &a_2P_{n_1,n_2}
    =
    \sqrt{
    \frac{
    n_2(n_2+\alpha)
    (N-n_2+1)
    (N-n_1-n_2+1)}
    {N^2}}
    P_{n_1,n_2-1},
    \nonumber\\
    &a_2^\dagger P_{n_1,n_2}
    =
    \sqrt{
    \frac{
    (n_2+1)(n_2+\alpha+1)
    (N-n_2)
    (N-n_1-n_2)}
    {N^2}}
    P_{n_1,n_2+1}.
    \label{eq:prop-eigenfct}
\end{align}
This completes the algebraic solution of the discrete
Smorodinsky--Winternitz II model. The spectrum, ladder-operator
structure and orthogonal eigenfunctions are thus obtained explicitly.

\section{Continuum limit}
\label{sec:continuum-limit}

One of the main objectives of the present construction is to show that
the finite model introduced above converges, under an appropriate
scaling, to the continuous Smorodinsky--Winternitz II system.
Throughout this section we therefore let the lattice size
$N\rightarrow\infty$ while simultaneously rescaling the lattice
coordinates so that both the Hamiltonian and its eigenfunctions admit
non-trivial differential limits.

The rescaled variables are given by
\begin{equation}
    x_1(x)=pN+\sqrt{2p(1-p)N}\,x,
    \qquad
    x_2(x,y)=N-x_1(x)-\frac{y}{p},
    \qquad
    p=N^{-\theta},
    \label{eq:change-of-variables}
\end{equation}
where $0<\theta<1/2$, and in the limit $N\rightarrow\infty$, one has
$x\in\mathbb R$ and $y\ge0$. Equivalently,
\begin{equation}
    x(x_1) = \frac{x_1-pN}{\sqrt{2p(1-p)N}}, \qquad y(x_1,x_2) = p(N-x_1-x_2).
\end{equation}
Under the elementary shifts
\[
(x_1,x_2)\mapsto (x_1+a,x_2+b),
\qquad a,b\in\mathbb{Z},
\]
the induced variations of the continuum variables $x(x_1)$ and $y(x_1,x_2)$ are respectively,
\begin{equation}
    \delta x(a)
    =
    \frac{a}{\sqrt{2p(1-p)N}},
    \qquad
    \delta y(a,b)
    =
    -p(a+b).
    \label{eq:delta-continuum}
\end{equation}
Thus, for a smooth test function \(f\), the finite difference operator \eqref{eq:difference-operator} can be rewritten as a Taylor expansion  around the induced variations $\delta x$ and $\delta y$:
\begin{equation}
    \Delta_{a,b} f(x_1,x_2)
    =
    f(x_1+a,x_2+b)-f(x_1,x_2)
    =
    f(x+\delta x,y+\delta y)-f(x,y)
    =
    \sum_{\substack{i,j\geq 0\\(i,j)\neq(0,0)}}
    \frac{\delta x^i}{i!}
    \frac{\delta y^j}{j!}
    \partial_x^i\partial_y^j f(x,y).
    \label{eq:taylor-difference}
\end{equation}

\subsection{Hamiltonian and eigenfunctions}

Applying the expansion \eqref{eq:taylor-difference} to the commuting number operators \eqref{eq:number-operators} and to the
ladder operators \eqref{eq:explicit-annihilation}-\eqref{eq:explicit-creation}, and letting
$N\rightarrow\infty$, yields
\begin{align}
    &\lim_{N\rightarrow\infty}N_1
    =
    -\frac12\partial_x^2+x\partial_x,
    \qquad
    \lim_{N\rightarrow\infty}N_2
    =
    -y\partial_y^2-(\alpha+1-y)\partial_y,
    \nonumber\\
    &
    \lim_{N\rightarrow\infty}a_1
    =
    \frac1{\sqrt2}\partial_x,
    \qquad
    \lim_{N\rightarrow\infty}a_2
    =
    -y\partial_y^2-(\alpha+1)\partial_y,
    \nonumber\\
    &
    \lim_{N\rightarrow\infty}a_1^\dagger
    =
    -\frac{\partial_x-2x}{\sqrt2},
    \qquad
    \lim_{N\rightarrow\infty}a_2^\dagger
    =
    -y\partial_y^2-(\alpha+1-2y)\partial_y+\alpha+1-y.
    \label{eq:limit-operators}
\end{align}

These limiting operators have an immediate interpretation.
The operators
$N_1$, $a_1$ and $a_1^\dagger$
become respectively the differential operator together with the forward
and backward shift operators for the (normalized) Hermite polynomials; see \eqref{eq:Hermite-diff} and \eqref{eq:Hermite-shift}.
Likewise,
$N_2$, $a_2$ and $a_2^\dagger$
become the corresponding operators for the (normalized) Laguerre polynomials; see \eqref{eq:Laguerre-diff} and \eqref{eq:Laguerre-shift}.
It is therefore natural to expect that the finite eigenfunctions
constructed in Section~\ref{sec:ladder} converge to products of Hermite and Laguerre
polynomials.

Indeed, using the limit relation
\eqref{eq:limit-Krawtchouk-Hermite}
for the modified Krawtchouk factor and, for the modified dual Hahn
factor, first applying the hypergeometric identity
\eqref{eq:identity-hypergeometric}
followed by the limit
\eqref{eq:limit-hypergeometric},
one obtains
\begin{align}
    &
    \lim_{N\rightarrow\infty}
    K_{n_1,n_2}(x_1(x);p,N)
    =
    \frac{1}{\sqrt{2^{n_1}n_1!}}
    H_{n_1}(x),
    \nonumber\\
    &
    \lim_{N\rightarrow\infty}
    R_{n_2}(x_1(x),x_2(x,y);\alpha,N)
    =
    \frac{1}
    {\sqrt{\Gamma(n_2+\alpha+1)/n_2!}}
    L_{n_2}^{(\alpha)}(y),
\end{align}
where
$H_n$
and
$L_n^{(\alpha)}$
denote the Hermite and Laguerre polynomials introduced in
Appendix~\ref{app:hyp-poly}. The limit \eqref{eq:limit-Krawtchouk-Hermite} is stated for fixed
$p$; it remains valid under the scaling $p=N^{-\theta}$ of \eqref{eq:change-of-variables} since
$Np(1-p)\to\infty$ for $\theta<1$. The vanishing of $p$ is in fact needed here: it is what makes
the factor $(1-p)^{-n_2/2}$ in \eqref{eq:mod-Krawtchouk} tend to one, so that the limit of
$K_{n_1,n_2}$ no longer depends on $n_2$.

The continuum limit of the weight function \eqref{eq:weight-function} is obtained by combining De Moivre--Laplace's theorem and Stirling's approximation with the scaling
\eqref{eq:change-of-variables}. One finds
\begin{equation}
    w_1(x_1(x)) \approx \frac{1}{\sqrt{2\pi p(1-p)N}}\,e^{-x^2}, \qquad w_2\big(x_1(x),x_2(x,y)\big) \approx py^\alpha e^{-y}.
\end{equation}
Moreover, the determinant of the Jacobian matrix yields
\begin{equation}
    \left|
    \frac{\partial(x_1,x_2)}{\partial(x,y)}
    \right|
    =
    \frac{\sqrt{2p(1-p)N}}{p}.
    \label{eq:jacobian-continuum}
\end{equation}
Consequently,
\begin{equation}
    \left|
    \frac{\partial(x_1,x_2)}{\partial(x,y)}
    \right|w(x_1(x),x_2(x,y)) \xrightarrow{N \to \infty}\frac{1}{\sqrt{\pi}}e^{-x^2}y^\alpha e^{-y}.
\end{equation}
In the continuum limit, the orthogonality relation
\eqref{eq:ortho-rel-energy-fct}
therefore converges to
\begin{equation}
    \frac{1}{\sqrt{\pi}}\int_{-\infty}^\infty e^{-x^2}\frac{H_{m_1}(x)H_{n_1}(x)}{\sqrt{2^{m_1}m_1!2^{n_1}n_1!}}dx\int_0^\infty y^\alpha e^{-y}\frac{L^{(\alpha)}_{m_2}(y)L^{(\alpha)}_{n_2}(y)}{\sqrt{\frac{\Gamma(m_2+\alpha+1)}{m_2!}\frac{\Gamma(n_2+\alpha+1)}{n_2!}}}dy = \delta_{m_1,n_1}\delta_{m_2,n_2}
    \label{eq:continuum-orthogonality}
\end{equation}
which is exactly the orthogonality relation for the product basis of
Hermite and Laguerre polynomials. As in the Smorodinsky--Winternitz I case \cite{Che-Vin-26}, to recover the Schrödinger representation of the continuous Smorodinsky--Winternitz II system, we now rescale
\begin{equation}
    y \mapsto y^2/2,
    \label{eq:y-to-ysquared}
\end{equation}
and introduce the gauge factor
\begin{equation}
    g(x,y) = \frac{1}{\pi^{1/4}}e^{-x^2/2}y^{\alpha+1/2}e^{-y^2/4}.
\end{equation}
Proceeding with the limiting operators \eqref{eq:limit-operators}, one obtains
\begin{align}
    &\cA_1 = g\left(\lim_{N \to \infty} a_1\right)g^{-1} = \tfrac{1}{\sqrt{2}}\left(\partial_x+x\right),
    \nonumber\\
    &\cA_1^\dagger = g\left(\lim_{N \to \infty} a_1^\dagger\right)g^{-1} = \tfrac{1}{\sqrt{2}}\left(-\partial_x+x\right),
    \nonumber\\
    &\cN_1 = g\left(\lim_{N \to \infty} N_1\right)g^{-1} = -\frac{1}{2}\partial_x^2+\frac{x^2}{2}-\frac{1}{2},
    \nonumber\\
    &\cA_2 = g\left(\left.\lim_{N\to\infty}a_2\right|_{y\mapsto y^2/2}\right)g^{-1} = -\tfrac{1}{2}\left(\partial_y^2+y\partial_y+\tfrac{y^2}{4}+\tfrac{1-4\alpha^2}{4y^2}+\tfrac{1}{2}\right),
    \nonumber\\
    &\cA_2^\dagger = g\left(\left.\lim_{N\to\infty}a_2^\dagger\right|_{y\mapsto y^2/2}\right)g^{-1} = -\tfrac{1}{2}\left(\partial_y^2-y\partial_y+\tfrac{y^2}{4}+\tfrac{1-4\alpha^2}{4y^2}-\tfrac{1}{2}\right),
    \nonumber\\
    &\cN_2 = g\left(\left.\lim_{N\to\infty}N_2\right|_{y\mapsto y^2/2}\right)g^{-1} = -\frac{1}{2}\partial_y^2+\frac{y^2}{8}-\frac{1-4\alpha^2}{8y^2}-\frac{\alpha+1}{2}.
\end{align}
It follows that
\begin{equation}
    \cH = g\left(\left.\lim_{N\to\infty}H\right|_{y\mapsto y^2/2}\right)g^{-1} = \cN_1 + \cN_2 + \tfrac{\alpha}{2} + 1 = -\frac{1}{2}\partial_x^2 - \frac{1}{2}\partial_y^2 + \frac{1}{8}\left(4x^2+y^2\right) - \frac{1-4\alpha^2}{8y^2}.
\end{equation}
This is the continuous Smorodinsky--Winternitz II system as defined in \cite{Mil-Pos-Win-13}. Its associated normalized eigenfunctions are \cite{Tem-Tur-Win-00}
\begin{align}
    \Psi_{n_1,n_2}(x,y) &= g(x,y)\lim_{N\to\infty}P_{n_1,n_2}\left(x_1(x),x_2\left(x,y^2/2\right)\right),
    \nonumber\\
    &= \frac{1}{\pi^{1/4}\sqrt{2^{n_1}n_1!}}e^{-x^2/2}H_{n_1}(x)\sqrt{\frac{n_2!}{\Gamma(\alpha+n_2+1)}}y^{\alpha+1/2}e^{-y^2/4}L_{n_2}^{(\alpha)}\left(y^2/2\right)
\end{align}
namely the familiar Hermite--Laguerre normalized eigenfunctions obtained by separation of variables in Cartesian and parabolic coordinates.

\subsection{Symmetry algebra}

Having recovered the continuous Smorodinsky--Winternitz II
Hamiltonian, it remains to determine the fate of its symmetry algebra.
An important question is whether the Hahn presentation of the discrete
symmetry algebra survives the continuum limit. We shall show that this
is not the case. While the symmetry operators themselves possess
regular limits, the Hahn presentation becomes singular and is replaced
by the Laguerre--Heun algebra naturally associated with the continuous
system.

Applying the scalings \eqref{eq:change-of-variables} and \eqref{eq:y-to-ysquared}, followed by the gauge transformation introduced above, to the symmetry operators \eqref{eq:symmetry-operators} of Section \ref{sec:ladder} yields
\begin{align}
    &\cC_1 = g\left(\left.\lim_{N\to\infty}C_1\right|_{y\mapsto y^2/2}\right)g^{-1} = -\frac{1}{2}\left(\partial_y^2 - \frac{y^2}{4} + \frac{1-4\alpha^2}{4y^2}\right),
    \nonumber\\
    &\cC_2 = g\left(\left.\lim_{N\to\infty}C_2\right|_{y\mapsto y^2/2}\right)g^{-1} = -\sqrt{2}\left(\frac{1}{2}\{(x\partial_y-y\partial_x),\partial_y\} + \frac{xy^2}{4} + \frac{(1-4\alpha^2)x}{4y^2}\right),
    \label{eq:limit-sym-op}
\end{align}
which are respectively, the complementary Cartesian and parabolic separation operators up to normalisation. One might expect the Hahn algebra obtained in \eqref{eq:hahn-algebra} to converge directly to the symmetry algebra of the continuous system. This expectation is, however, incorrect, and it is instructive to track the structure constants \eqref{eq:struc-const} under the scaling \eqref{eq:change-of-variables}.

Since the Hamiltonian eigenvalue stays finite while $N\rightarrow\infty$, one has $\eta \sim 4/N$, so that the structure constants tend to finite limits,
\begin{equation}
    \mu\longrightarrow 0,
    \qquad
    \nu\longrightarrow -24,
    \qquad
    \xi\longrightarrow 16\,\mathcal H,
    \qquad
    \zeta\longrightarrow 2(\alpha^2-1).
    \label{eq:limit-structure-constants}
\end{equation}
The cubic term thus drops out and the cubic presentation degenerates
regularly to a quadratic algebra. By contrast, the coefficients
of the Hahn presentation \eqref{eq:hahn-pres} become singular:
\begin{equation}
    \kappa=\sqrt{\tfrac{\mu}{2}}=2\sqrt{\eta}\sim\frac{4}{\sqrt N}\longrightarrow 0,
    \qquad
    \lambda=\frac{\nu}{3\kappa}\sim-2\sqrt N\longrightarrow-\infty,
    \qquad
    \xi-\lambda^2\longrightarrow-\infty .
    \label{eq:limit-hahn-coefficients}
\end{equation}
Hence the redefinition $\widetilde C_2=C_2+\kappa C_1^2+\lambda C_1$ and
the Hahn structure constants diverge, and the Hahn presentation does not
possess a meaningful limit.

The singular behaviour originates from the normalization adapted to the
finite lattice. Although this normalization provides a natural
presentation of the discrete symmetry algebra, it is incompatible with
the continuum scaling. The symmetry operators themselves admit regular
limits, but the polynomial relations satisfied by the renormalized
generators degenerate. It is the vanishing of $\eta$ that is responsible: in the companion
model \cite{Che-Vin-26} the corresponding structure function tends instead to a finite non-zero value, and the Hahn presentation there survives the limit unscathed.

To obtain the correct limiting algebra one retains instead the
differential realizations \eqref{eq:limit-sym-op} of
the limiting symmetry operators. A direct computation gives the quadratic algebra
\begin{align}
    [\mathcal C_1,
     \mathcal C_2]
    &\equiv
    \mathcal C_3,
    \nonumber\\
    [\mathcal C_2,
     \mathcal C_3]
    &=
    -24\,\mathcal C_1^{\,2}
    +16\,\mathcal H\,\mathcal C_1
    +2(\alpha^2-1),
    \nonumber\\
    [\mathcal C_3,
     \mathcal C_1]
    &=
    -\mathcal C_2,
    \label{eq:laguerre-heun-algebra}
\end{align}
with $\mathcal H$ central. These relations are exactly the limits
\eqref{eq:limit-structure-constants} of the discrete cubic algebra: the
$\mathcal C_1^3$ term has disappeared, and the surviving
coefficients  are precisely the limiting structure constants \eqref{eq:limit-structure-constants}. The algebra
\eqref{eq:laguerre-heun-algebra} is the Laguerre--Heun algebra of the
continuous Smorodinsky--Winternitz II system identified
in~\cite{chea20262d}, with $\mathcal C_1$ the Laguerre
operator, $\mathcal C_2$ its Heun partner and
$\mathcal C_3$ their commutator; the structure constants here
match those of~\cite{chea20262d} up to the normalization of
$\mathcal C_1$ and $\mathcal C_2$.

The continuum limit therefore reveals a remarkable phenomenon.
Although the discrete symmetry algebra admits a Hahn-algebra
presentation, that presentation is not stable under the continuum
limit. Instead, the limiting symmetry operators close according to the
Laguerre--Heun algebra naturally associated with the
Smorodinsky--Winternitz II system.
This observation provides a direct algebraic link between the finite
model constructed in the present paper and the continuous theory
developed in our recent work on the Laguerre--Heun
algebra~\cite{chea20262d}. In this
sense, the discrete model furnishes a finite realization whose
continuum limit explains the emergence of the Laguerre--Heun symmetry
algebra from a finite Hahn-algebra framework.
The continuum limit therefore preserves not only the Hamiltonian and
its eigenfunctions but also the underlying superintegrable structure,
although its most natural algebraic presentation changes in passing
from the finite lattice to the continuous setting.

\section{Conclusion}
\label{sec:conclusion}

We have constructed a finite discrete analogue of the
Smorodinsky--Winternitz II superintegrable system on a triangular region of the two-dimensional square lattice. The model is defined in terms of two commuting finite-difference number operators together with an associated ladder-operator structure. We have shown that it is maximally
superintegrable and that its symmetry algebra admits a Hahn-algebra
presentation. The spectral problem has been solved exactly in terms of
the bivariate Krawtchouk and dual Hahn polynomials of Tratnik type associated with the factorized $A_2$-Leonard pair.

An important feature of the construction is that it provides a genuine
finite realization of the continuous Smorodinsky--Winternitz II system rather than merely a finite-difference approximation of its equations of motion. Under an appropriate
continuum limit, the Hamiltonian, the ladder operators and the
eigenfunctions converge respectively to their continuous counterparts,
recovering the Hermite--Laguerre separation of variables together with
the associated Schrödinger presentation.

The continuum limit also reveals an interesting algebraic phenomenon.
Although the discrete symmetry algebra naturally admits a Hahn-algebra
presentation, this presentation becomes singular under the continuum
scaling. The limiting symmetry operators nevertheless possess regular
differential realizations and close according to the
Laguerre--Heun algebra associated with the continuous
Smorodinsky--Winternitz II system. The present construction therefore
provides a natural discrete origin for this algebra and clarifies how it
emerges naturally from a finite superintegrable model.

Together with our companion paper devoted to the
Smorodinsky--Winternitz I system \cite{Che-Vin-26}, the present work provides further
evidence that finite discrete models can preserve the essential
features of superintegrability, including exact solvability, ladder
structures, symmetry algebras and continuum limits. We hope that these
results will stimulate further investigations of finite
superintegrable systems associated with other families of multivariate
orthogonal polynomials and their underlying algebraic structures.

\section*{Acknowledgments}
PAB acknowledges support from a CQIQC postdoctoral fellowship and a postdoctoral fellowship
from the Fonds de Recherche du Québec – Nature et Technologies (FRQNT). VVC benefits from a scholarship from the FRQNT. The work of LV is supported in part through a Natural Sciences and Engineering Research Council (NSERC) of Canada.

\section*{Conflict of interest}
The authors state that there is no conflict of interest.

\section*{Data availability}
This manuscript has no associated data.

\appendix

\section{Coefficients of the ladder operators}
\label{app:ladder-coefficients}

The coefficients of the annihilation operators~\eqref{eq:explicit-annihilation} are given by
\begin{align}
    &S_1 = \{(1,0),(0,1),(1,-1),(-1,1),(-1,0),(0,-1)\},
    \nonumber\\
    &a_{1}^{1,0} = \frac{p \left(N-x_1-x_2\right) \left(\alpha +2 N-x_1-x_2+1\right)}{\alpha +2 N-x_1-2 x_2+1},
    \nonumber\\
    &a_{1}^{0,1} = -\frac{p \left(N-x_2\right) \left(N-x_1-x_2\right) \left(\alpha +2 N-x_1-x_2+1\right)}{\left(\alpha +2 N-x_1-2 x_2\right)_2},
    \nonumber\\
    &a_{1}^{1,-1} = \frac{p x_2 \left(\alpha +N-x_2+1\right)}{\alpha +2 N-x_1-2 x_2+1},
    \qquad
    a_{1}^{-1,1} = -\frac{p x_1 \left(N-x_2\right)}{\alpha +2 N-x_1-2 x_2+1},
    \nonumber\\
    &a_{1}^{-1,0} = -\frac{p x_1 \left(\alpha +N-x_1-x_2+1\right)}{\alpha +2 N-x_1-2 x_2+1}, \qquad
    a_{1}^{0,-1} = -\frac{p x_2 \left(\alpha +N-x_2+1\right) \left(\alpha +N-x_1-x_2+1\right)}{\left(\alpha +2 N-x_1-2 x_2+1\right)_2},
    \label{eq:a1-coefficients}
\end{align}
\begin{align}
    S_2 = \ & \{(0,2),(-1,2),(0,1),(-1,1),(-1,0),(0,-1),(-1,-1),(0,-2)\},
    \nonumber\\
    a_{2}^{0,2} = \ &-\frac{\left(N-x_2-1\right)_2 \left(N-x_1-x_2-1\right)_2 \left(\alpha +2 N-x_1-x_2\right)_2}{\left(\alpha +2 N-x_1-2 x_2-2\right)_4} ,
    \nonumber\\
    a_{2}^{-1,2} = \ &-\frac{x_1 \left(N-x_2-1\right)_2 \left(N-x_1-x_2\right) \left(\alpha +2 N-x_1-x_2+1\right)}{\left(\alpha +2 N-x_1-2
   x_2-1\right)_3},
    \nonumber\\
    a_{2}^{0,1} = \ &-\frac{\left(x_2+2\right) \left(N-x_1-x_2\right) \left(N-x_2\right){}^2 \left(\alpha +N-x_2\right) \left(\alpha +2
   N-x_1-x_2+1\right)}{\left(\alpha +2 N-x_1-2 x_2-1\right)_3 \left(\alpha +2 N-x_1-2 x_2\right){}}
    \nonumber\\
    &+\frac{\left(N-x_2\right)_2 \left(N-x_1-x_2\right)_2 \left(\alpha +2
   N-x_1-x_2+1\right)_2}{\left(\alpha +2 N-x_1-2 x_2\right)_3 \left(\alpha +2 N-x_1-2 x_2+1\right){}}
    \nonumber\\
    &+\frac{\left(N-x_1-x_2\right) \left(N-x_2\right) \left(\alpha +N-x_2\right) \left(\alpha
   +N-x_1-x_2\right) \left(\alpha +2 N-x_1-x_2+1\right)_2}{\left(\left(\alpha +2 N-x_1-2 x_2\right){}_2\right)^2}
    \nonumber\\
    &-\frac{\left(x_2+2\right) \left(N-x_2\right) \left(N-x_1-x_2-1\right)_2
   \left(\alpha +N-x_1-x_2-1\right) \left(\alpha +2 N-x_1-x_2+1\right)}{\left(\alpha +2 N-x_1-2 x_2-2\right)_4},
    \nonumber\\
    a_{2}^{-1,1} = \ &\frac{x_1 \left(N-x_2\right)_2 \left(N-x_1-x_2+1\right) \left(\alpha +2 N-x_1-x_2+2\right)}{\left(\alpha +2 N-x_1-2
   x_2+1\right){}^2 \left(\alpha +2 N-x_1-2 x_2+2\right)}
    \nonumber\\
    &+\frac{x_1 \left(N-x_2\right) \left(\alpha +N-x_2\right) \left(\alpha
   +N-x_1-x_2\right) \left(\alpha +2 N-x_1-x_2+2\right)}{\left(\alpha +2 N-x_1-2 x_2\right) \left(\alpha +2 N-x_1-2
   x_2+1\right){}^2}
    \nonumber\\
    &-\frac{x_1 \left(x_2+2\right) \left(N-x_2\right) \left(N-x_1-x_2\right) \left(\alpha +N-x_1-x_2\right)}{\left(\alpha +2
   N-x_1-2 x_2-1\right)_3},
    \nonumber\\
    a_{2}^{-1,0} = \ &-\frac{x_1 \left(N-x_2+1\right) \left(\alpha +N-x_2+1\right) \left(\alpha +N-x_1-x_2+1\right) \left(\alpha +2 N-x_1-x_2+3\right)}{\left(\alpha
   +2 N-x_1-2 x_2+1\right)_3}
    \nonumber\\
    &+\frac{x_1 \left(x_2+1\right)
   \left(N-x_2+1\right) \left(N-x_1-x_2+1\right) \left(\alpha +N-x_1-x_2+1\right)}{\left(\alpha +2 N-x_1-2 x_2+1\right){}^2 \left(\alpha +2
   N-x_1-2 x_2+2\right)}
    \nonumber\\
    &+\frac{x_1 \left(x_2+1\right) \left(\alpha +N-x_2\right) \left(\alpha +N-x_1-x_2\right) \left(\alpha
   +N-x_1-x_2+1\right)}{\left(\alpha +2 N-x_1-2 x_2\right) \left(\alpha +2 N-x_1-2 x_2+1\right){}^2},
    \nonumber\\
    a_{2}^{0,-1} = \ &-\frac{x_2 \left(N-x_1-x_2+1\right) \left(\alpha +N-x_2+1\right) \left(\alpha +2 N-x_1-x_2+3\right) \left(\alpha
   +N-x_1-x_2+1\right){}^2}{\left(\alpha +2 N-x_1-2 x_2+1\right)_3 \left(\alpha +2 N-x_1-2 x_2+2\right){}}
    \nonumber\\
    &-\frac{x_2 \left(N-x_2+2\right) \left(\alpha +N-x_2+1\right)_2 \left(\alpha +2 N-x_1-x_2+3\right)
   \left(\alpha +N-x_1-x_2+1\right)}{\left(\alpha +2 N-x_1-2 x_2+1\right)_4}
    \nonumber\\
    &+\frac{(x_2)_2 \left(N-x_2+1\right) \left(N-x_1-x_2+1\right) \left(\alpha
   +N-x_2+1\right) \left(\alpha +N-x_1-x_2+1\right)}{\left(\left(\alpha +2 N-x_1-2 x_2+1\right){}_2\right)^2}
    \nonumber\\
    &+\frac{(x_2)_2 \left(\alpha +N-x_2\right)_2 \left(\alpha +N-x_1-x_2\right)_2}{\left(\alpha +2 N-x_1-2 x_2\right)_3 \left(\alpha +2 N-x_1-2 x_2+1\right){}},
    \nonumber\\
    a_{2}^{-1,-1} = \ &-\frac{x_1 x_2 \left(\alpha +N-x_2+1\right) \left(\alpha +N-x_1-x_2+1\right)_2}{\left(\alpha +2 N-x_1-2
   x_2+1\right)_3},
    \nonumber\\
    a_{2}^{0,-2} = \ &-\frac{\left(x_2-1\right)_2 \left(\alpha +N-x_2+1\right)_2  \left(\alpha +N-x_1-x_2+1\right)_2}{\left(\alpha +2 N-x_1-2 x_2+1\right)_4}.
    \label{eq:a2-coefficients}
\end{align}
The coefficients of the creation operators~\eqref{eq:explicit-creation} are given by
\begin{align}
    &\bar{S}_1 = \{(1,0),(0,1),(1,-1),(0,0),(-1,1),(-1,0),(0,-1)\},
    \nonumber\\
    &\bar a_{1}^{1,0} =
    -\frac{p^2 \left(N-x_1-x_2\right) \left(\alpha +2 N-x_1-x_2+1\right)}{(1-p) \left(\alpha +2 N-x_1-2 x_2+1\right)},
    \nonumber\\
    &\bar a_{1}^{0,1} = -\frac{p \left(N-x_2\right) \left(N-x_1-x_2\right) \left(\alpha +2 N-x_1-x_2+1\right)}{\left(\alpha +2 N-x_1-2 x_2\right)_2},
    \nonumber\\
    &\bar a_{1}^{1,-1} = -\frac{p^2 x_2 \left(\alpha +N-x_2+1\right)}{(1-p) \left(\alpha +2 N-x_1-2 x_2+1\right)},
    \qquad
    \bar a_{1}^{0,0} = -\frac{N p-x_1}{1-p},
    \qquad
    \bar a_{1}^{-1,1} = \frac{(1-p) x_1 \left(N-x_2\right)}{\alpha +2 N-x_1-2 x_2+1},
    \nonumber\\
    &\bar a_{1}^{-1,0} = \frac{(1-p) x_1 \left(\alpha +N-x_1-x_2+1\right)}{\alpha +2 N-x_1-2 x_2+1}, \qquad
    \bar a_{1}^{0,-1} = -\frac{p x_2 \left(\alpha +N-x_2+1\right) \left(\alpha +N-x_1-x_2+1\right)}{\left(\alpha +2 N-x_1-2 x_2+1\right)_2},
    \label{eq:a1d-coefficients}
\end{align}
\begin{align}
    \bar{S}_2 =&\ \{(0,2),(1,1),(1,0),(0,1),(0,0),(1,-1),(1,-2),(0,-1),(0,-2)\},
    \nonumber\\
    \bar a_{2}^{0,2} =&\
    -\frac{\left(N-x_2-1\right)_2 \left(N-x_1-x_2-1\right)_2 \left(\alpha +2 N-x_1-x_2\right)_2}{\left(\alpha +2 N-x_1-2 x_2-2\right)_4},
    \nonumber\\
    \bar a_{2}^{1,1} =&\
    -\frac{p \left(N-x_2\right) \left(N-x_1-x_2-1\right)_2 \left(\alpha +2 N-x_1-x_2\right)_2}{(1-p) \left(\alpha +2 N-x_1-2 x_2-1\right)_3},
    \nonumber\\
    \bar a_{2}^{1,0} =&\
    \frac{p x_2 \left(N-x_2\right) \left(N-x_1-x_2\right){}^2 \left(\alpha +2 N-x_1-x_2+1\right)}{(1-p) \left(\alpha +2 N-x_1-2 x_2\right)
   \left(\alpha +2 N-x_1-2 x_2+1\right){}^2}
    \nonumber\\
    &-\frac{p \left(N-x_2\right) \left(N-x_1-x_2\right) \left(\alpha +N-x_2\right) \left(\alpha +2
   N-x_1-x_2\right)_2}{(1-p) \left(\alpha +2 N-x_1-2 x_2-1\right)_3}
    \nonumber\\
    &+\frac{p x_2 \left(N-x_1-x_2\right) \left(\alpha +N-x_2+1\right) \left(\alpha +N-x_1-x_2+1\right)
   \left(\alpha +2 N-x_1-x_2+1\right)}{(1-p) \left(\alpha +2 N-x_1-2 x_2+1\right){}^2 \left(\alpha +2 N-x_1-2 x_2+2\right)},
    \nonumber\\
    \bar a_{2}^{0,1} =&\
    -\frac{\left(N-x_2\right) \left(N-x_1-x_2\right){}^2 \left(\alpha +N-x_1-x_2\right) \left(\alpha +2 N-x_1-x_2\right)_2}{\left(\alpha +2 N-x_1-2 x_2-1\right)_3 \left(\alpha +2 N-x_1-2 x_2\right){}}
    \nonumber\\
    &+\frac{x_2 \left(N-x_2\right){}^2 \left(N-x_1-x_2\right){}^2 \left(\alpha +2 N-x_1-x_2+1\right)}{\left(\left(\alpha +2 N-x_1-2
   x_2\right){}_2\right)^2}
    \nonumber\\
    &-\frac{\left(N-x_2-1\right)_2 \left(N-x_1-x_2\right) \left(\alpha
   +N-x_2-1\right) \left(\alpha +2 N-x_1-x_2\right)_2}{\left(\alpha +2 N-x_1-2 x_2-2\right)_4}
    \nonumber\\
    &+\frac{x_2 \left(N-x_2\right)
   \left(N-x_1-x_2\right) \left(\alpha +N-x_2+1\right) \left(\alpha +N-x_1-x_2+1\right) \left(\alpha +2 N-x_1-x_2+1\right)}{\left(\alpha +2
   N-x_1-2 x_2\right)_3 \left(\alpha +2 N-x_1-2 x_2+1\right){}},
    \nonumber\\
    \bar a_{2}^{0,0} =&\ -\frac{\left(N-x_2\right) \left(N-x_1-x_2\right)-(\alpha +1) x_2}{1-p},
    \nonumber\\
    \bar a_{2}^{1,-1} =&\ \frac{p x_2 \left(\alpha +N-x_1-x_2+1\right) \left(\alpha +2 N-x_1-x_2+1\right) \left(\alpha +N-x_2+1\right){}^2}{(1-p) \left(\alpha +2
   N-x_1-2 x_2+1\right){}^2 \left(\alpha +2 N-x_1-2 x_2+2\right)}
    \nonumber\\
    &+\frac{p x_2 \left(N-x_2\right) \left(N-x_1-x_2\right) \left(\alpha +2
   N-x_1-x_2+1\right) \left(\alpha +N-x_2+1\right)}{(1-p) \left(\alpha +2 N-x_1-2 x_2\right) \left(\alpha +2 N-x_1-2 x_2+1\right){}^2}
    \nonumber\\
    &-\frac{p
   \left(x_2-1\right) x_2 \left(N-x_1-x_2+1\right) \left(\alpha +N-x_1-x_2+1\right) \left(\alpha +N-x_2+1\right)}{(1-p) \left(\alpha +2
   N-x_1-2 x_2+1\right)_3},
    \nonumber\\
    \bar a_{2}^{0,-1} =&\ \frac{x_2 \left(\alpha +N-x_1-x_2+1\right){}^2 \left(\alpha +2 N-x_1-x_2+1\right) \left(\alpha +N-x_2+1\right){}^2}{\left(\left(\alpha +2 N-x_1-2
   x_2+1\right){}_2\right)^2}
    \nonumber\\
    &-\frac{\left(x_2-1\right)_2 \left(N-x_2+1\right) \left(\alpha
   +N-x_1-x_2+1\right) \left(\alpha +N-x_2+1\right){}^2}{\left(\alpha +2 N-x_1-2 x_2+1\right)_3 \left(\alpha +2 N-x_1-2 x_2+2\right){}}
    \nonumber\\
    &+\frac{x_2 \left(N-x_2\right) \left(N-x_1-x_2\right) \left(\alpha +N-x_1-x_2+1\right) \left(\alpha +2
   N-x_1-x_2+1\right) \left(\alpha +N-x_2+1\right)}{\left(\alpha +2 N-x_1-2 x_2\right)_3 \left(\alpha +2 N-x_1-2 x_2+1\right){}}
    \nonumber\\
    &-\frac{\left(x_2-1\right)_2 \left(N-x_1-x_2+2\right) \left(\alpha +N-x_1-x_2+1\right)_2 \left(\alpha +N-x_2+1\right)}{\left(\alpha +2 N-x_1-2 x_2+1\right)_4},
    \nonumber\\
    \bar a_{2}^{1,-2} =&\
    -\frac{p \left(x_2-1\right)_2 \left(\alpha +N-x_2+1\right)_2 \left(\alpha +N-x_1-x_2+1\right)}{(1-p) \left(\alpha
   +2 N-x_1-2 x_2+1\right)_3}
    \nonumber\\
    \bar a_{2}^{0,-2} =&\
    -\frac{\left(x_2-1\right)_2 \left(\alpha +N-x_2+1\right)_2 \left(\alpha +N-x_1-x_2+1\right)_2}{\left(\alpha +2 N-x_1-2 x_2+1\right)_4}.
    \label{eq:a2d-coefficients}
\end{align}

\section{Hypergeometric polynomials}
\label{app:hyp-poly}

\subsection{Hypergeometric function}

The hypergeometric function is defined by \cite{Koe-Les-Swa-10}
\begin{equation}
    _r{F}_s\left(\genfrac{}{}{0pt}{}{a_1,\dotsc, a_r}{b_1,\dotsc,b_s};z\right) = \sum_{k=0}^\infty \frac{(a_1)_k\cdots(a_r)_k}{(b_1)_k\cdots(b_s)_k}\frac{z^k}{k!}.
    \label{eq:hypergeometric-function}
\end{equation}
They have the limit relations
\begin{align}
    &\lim_{\lambda\to\infty} \ _r{F}_s\left(\genfrac{}{}{0pt}{}{a_1,\dotsc, a_{r-1},\lambda a_r}{b_1,\dotsc,b_s};\frac{z}{\lambda}\right) = \ _{r-1}{F}_s\left(\genfrac{}{}{0pt}{}{a_1,\dotsc, a_{r-1}}{b_1,\dotsc,b_s};a_rz\right),
    \nonumber\\
    &\lim_{\lambda\to\infty} \ _r{F}_s\left(\genfrac{}{}{0pt}{}{a_1,\dotsc,a_r}{b_1,\dotsc,b_{s-1},\lambda b_s};\lambda z\right) = \ _r{F}_{s-1}\left(\genfrac{}{}{0pt}{}{a_1,\dotsc, a_r}{b_1,\dotsc,b_{s-1}};\frac{z}{b_s}\right),
    \label{eq:limit-hypergeometric}
\end{align}
which generalizes to
\begin{align}
    \lim_{\lambda\to\infty} &\ _r{F}_s\left({a_1,\dotsc,a_{r-i},\lambda^{k_1}a_{r-i+1},\dotsc,\lambda^{k_i} a_r\atop b_1,\dotsc,b_{s-i},\lambda^{l_1}b_{s-j+1},\dotsc,\lambda^{l_j}b_s};\frac{\lambda^{l_1}\cdots\lambda^{l_j}}{\lambda^{k_1}\cdots\lambda^{k_i}}z\right)
    \nonumber\\
    &= \ _{r-i}{F}_{s-j}\left({a_1,\dotsc, a_{r-i}\atop b_1,\dotsc,b_{s-j}};\frac{a_{r-i+1}\cdots a_r}{b_{s-j+1}\cdots b_s}z\right).
    \label{eq:general-limit-hypergeometric}
\end{align}
A specific functional identity for $_3{F}_2$ is given by
\begin{align}
    &_3{F}_2\left({a_1,a_2,a_3\atop b_1,b_2};1\right) = \frac{\Gamma(b_1)\Gamma(b_1+b_2-a_1-a_2-a_3)}{\Gamma(b_1-a_1)\Gamma(b_1+b_2-a_2-a_3)} \ _3{F}_2\left({a_1,b_2-a_2,b_2-a_3\atop b_1+b_2-a_2-a_3,b_2};1\right),
    \label{eq:identity-hypergeometric}
\end{align}
where $\text{Re}(b_1+b_2-a_1-a_2-a_3),\text{Re}(b_1-a_1)>0$ and $\Gamma(z) = \int_0^\infty t^{z-1}e^{-t}dt$, $\text{Re}(z)>0$, is the Gamma function.

\subsection{Formulas for the Hermite polynomials}

The Hermite polynomials are defined by \cite{Koe-Les-Swa-10}
\begin{equation}
    H_n(x) = (2x)^n\ _2{F}_0\left({-\frac{n}{2},-\frac{n-1}{2}\atop -};-\frac{1}{x^2}\right).
    \label{eq:Hermite}
\end{equation}
They satisfy the differential equation
\begin{equation}
    H^{''}_n(x) - 2xH^{'}_n(x) + 2nH_n(x) = 0.
    \label{eq:Hermite-diff}
\end{equation}
They obey the orthogonality relation
\begin{equation}
   \frac{1}{\sqrt{\pi}}\int_{-\infty}^\infty dx \ e^{-x^2}H_m(x)H_n(x) = 2^nn!\delta_{mn}.
   \label{eq:Hermite-orthogonality}
\end{equation}
The forward and backward shift operators are respectively as follow
\begin{align}
    &H^{'}_n(x) = 2nH_{n-1}(x),
    \nonumber\\
    &H^{'}_n(x) - 2xH_n(x) = - H_{n+1}(x).
    \label{eq:Hermite-shift}
\end{align}

\subsection{Formulas for the Laguerre polynomials}

The Laguerre polynomials are defined by \cite{Koe-Les-Swa-10}
\begin{equation}
    L_n^{(\alpha)}(x) = \frac{(\alpha+1)_n}{n!}\ _1{F}_1\left({-n\atop \alpha+1};x\right).
    \label{eq:Laguerre}
\end{equation}
They satisfy the differential equation
\begin{equation}
    xL^{(\alpha)''}_n(x) + (\alpha+1-x)L^{(\alpha)'}_n(x) + nL^{(\alpha)}_n(x) = 0.
    \label{eq:Laguerre-diff}
\end{equation}
They obey the orthogonality relation
\begin{equation}
    \int_0^\infty dx \ e^{-x}x^\alpha L_m^{(\alpha)}(x)L_n^{(\alpha)}(x) = \frac{\Gamma(n+\alpha+1)}{n!}\delta_{mn}, \quad \alpha>-1.
    \label{eq:Laguerre-orthogonality}
\end{equation}
The forward and backward shift operators are respectively as follow
\begin{align}
    &-xL^{(\alpha)''}_n(x) - (\alpha+1)L^{(\alpha)'}_n(x) = (n+\alpha)L^{(\alpha)}_{n-1}(x),
    \nonumber\\
    &-xL^{(\alpha)''}_n(x) - (\alpha+1-2x)L^{(\alpha)'}_n(x) + (\alpha+1-x)L^{(\alpha)}_n(x) = (n+1)L^{(\alpha)}_{n+1}(x).
    \label{eq:Laguerre-shift}
\end{align}

\subsection{Formulas for the Krawtchouk polynomials}

The Krawtchouk polynomials are defined by \cite{Koe-Les-Swa-10}
\begin{align}
    &\hat{K}_n(x;p,N) = \ _2{F}_1\left({-n,-x\atop -N};\frac{1}{p}\right), \quad n,x = 0,1,\dotsc,N, \quad 0<p<1.
    \label{eq:Krawtchouk}
\end{align}
They are related to the Hermite polynomials by the limit
\begin{align}
    &\lim_{N \to\infty}\sqrt{\binom{N}{n}}\hat{K}_n\left(pN+\sqrt{2p(1-p)N}x;p,N\right) = \frac{(-1)^nH_n(x)}{\sqrt{2^nn!\left(\frac{p}{1-p}\right)^n}}.
    \label{eq:limit-Krawtchouk-Hermite}
\end{align}

\subsection{Formulas for the dual Hahn polynomials}

The dual Hahn polynomials are defined by \cite{Koe-Les-Swa-10}
\begin{equation}
    \hat{R}_n(\lambda(x);\gamma,\delta,N) = \ _3{F}_2\left({-n,-x,x+\gamma+\delta+1\atop \gamma+1,-N};1\right), \qquad \lambda(x) = x(x+\gamma+\delta+1),
    \label{eq:dual-Hahn}
\end{equation}
where $N \in \mathbb{N}$, $n = 0,1,\dotsc,N$ and $\gamma,\delta>-1$ or $\gamma,\delta<-N$.

\section{Polynomials of Tratnik type}
\label{app:Tratnik-poly}

\subsection{Krawtchouk/dual Hahn}

The bivariate polynomials of the Krawtchouk and dual Hahn kinds \cite{tratnik1991some} are given by \cite{Cra-Zai-25}
\begin{equation}
    T_{i,j}(x,y) = \left(\frac{r}{1-r}\right)^i\binom{N-j}{i}\hat{K}_i(x;r,N-j)\frac{(-1)^j (\rho +1)_j}{(-N+\rho -\sigma )_j}\binom{N-x}{j}\hat{R}_j(\lambda(y);\rho,x-\rho+\sigma,N-x).
\end{equation}
where $\hat{K}_{i}$ and $\hat{R}_{j}$ are, respectively, the Krawtchouk polynomials \eqref{eq:Krawtchouk} and the dual Hahn polynomials \eqref{eq:dual-Hahn}.

They satisfy the difference equations
\begin{align}
    &i\,T_{i,j}(x,y) = D_{x+1,y}^{-1,0}\,T_{i,j}(x+1,y) + D_{x+1,y-1}^{-1,1}\,T_{i,j}(x+1,y-1) + D_{x-1,y+1}^{1,-1}\,T_{i,j}(x-1,y+1)
    \nonumber\\
    &\qquad \qquad \quad \ + D_{x,y}^{0,0}\,T_{i,j}(x,y) + D_{x-1,y}^{1,0}\,T_{i,j}(x-1,y) + D_{x,y-1}^{0,1}\,T_{i,j}(x,y-1),
    \nonumber\\
    &j\,T_{i,j}(x,y) = b_y\,T_{i,j}(x,y+1) + a_y\,T_{i,j}(x,y) + c_y\,T_{i,j}(x,y-1),
\end{align}
where the difference coefficients read as follow,
\begin{align}
    &D_{x+1,y}^{-1,0} = -\frac{r (N-x-y) (\sigma +x+y+1)}{\sigma +x+2 y+1}, \qquad D_{x+1,y-1}^{-1,1} = -\frac{r y (N+\sigma +y+1)}{\sigma +x+2 y+1},
    \nonumber\\
    &D_{x-1,y+1}^{1,-1} = \frac{(r-1) x (\rho +y+1)}{\sigma +x+2 y+1}, \qquad D_{x-1,y}^{1,0} = \frac{(r-1) x (-\rho +\sigma +x+y)}{\sigma +x+2 y+1},
    \nonumber\\
    &D_{x,y-1}^{0,1} = \frac{r y (N+\sigma +y+1) (-\rho +\sigma +x+y)}{(\sigma +x+2 y) (\sigma +x+2 y+1)},
    \nonumber\\
    &D_{x,y}^{0,0} = -\left(D_{x+1,y}^{-1,0} + D_{x+1,y-1}^{-1,1} + D_{x-1,y+1}^{1,-1} + D_{x-1,y}^{1,0} + D_{x,y-1}^{0,1}\right),
    \nonumber\\
    &b_y = \frac{(\rho +y+1) (-N+x+y) (\sigma +x+y+1)}{(\sigma +x+2 y+1) (\sigma +x+2 y+2)}, \qquad c_y = -\frac{y (N+\sigma +y+1) (-\rho +\sigma +x+y)}{(\sigma +x+2 y) (\sigma +x+2 y+1)},
    \nonumber\\
    &a_y = -(b_y + c_y).
\end{align}
They obey the orthogonality relation
\begin{equation}
    \sum_{x=0}^{\min(N-j,N-j')}\sum_{y=0}^{N-x}F(x,y)T_{i,j}(x,y)T_{i',j'}(x,y) = \delta_{i,i'}\delta_{j,j'}G(i,j)
\end{equation}
where the weight and the normalization functions are respectively given by
\begin{align}
    &F(x,y) = \binom{N}{x}r^x(1-r)^{N-x}(-1)^{N-x-y}\binom{N-x}{y}\frac{(\rho +1)_y (-N+\rho -\sigma )_{N-x-y}}{(x+y+\sigma +1)_y (x+2 y+\sigma
    +2)_{N-x-y}},
    \nonumber\\
    &G(i,j) = \left(\frac{r}{1-r}\right)^i \binom{N-j}{i} (-1)^j(1-r)^j\binom{N}{j}\frac{(\rho +1)_j}{(-N+\rho -\sigma )_j}.
\end{align}
Under the following change of variables,
\begin{align}
    &i \mapsto n_1, \qquad j \mapsto n_2, \qquad x \mapsto x_1, \qquad y \mapsto x_2
    \nonumber\\
    &r \mapsto p, \qquad \rho \mapsto -N-1, \qquad \sigma \mapsto -2N-\alpha-2,
\end{align}
we recover the bivariate polynomials \eqref{eq:energy-eigenfct} and the weight function \eqref{eq:weight-function},
\begin{align}
    &(-1)^{i+j}\frac{T_{i,j}(x,y)}{\sqrt{G(i,j)}} \mapsto \sqrt{\Gamma(\alpha+1)}P_{n_1,n_2}(x_1,x_2), \qquad F(x,y) \mapsto \frac{w(x_1,x_2)}{\Gamma(\alpha+1)}.
\end{align}

\bibliography{SWIIDR}

\end{document}